\documentclass[12pt,epsfig]{article}
\usepackage{amssymb,amsmath}
\usepackage{graphicx,epsfig}

\newcommand{\del}{\partial}

\newcommand{\be}{\begin{equation}}
\newcommand{\ee}{\end{equation}}
\newcommand{\bea}{\begin{eqnarray}}
\newcommand{\eea}{\end{eqnarray}}
\newcommand{\beas}{\begin{eqnarray*}}
\newcommand{\eeas}{\end{eqnarray*}}
\newcommand{\ba}{\begin{array}}
\newcommand{\ea}{\end{array}}

\newcommand{\is}{\! &\! = \! & \!}

\newcommand{\xX}{\mbox{\small {X}}}

\newpage

\newcommand{\smp}{\hspace{.5pt}}

\renewcommand{\sp}{\hspace{1pt}}

\newcommand{\dslash}{\partial \hspace{-6pt} /\sp}
\newcommand{\Dslash}{D\hspace{-7pt} /\sp}

\newcommand{\ccc}{\mbox{$c$\smp}}

\newcommand{\nbox}{{\,\lower0.9pt\vbox{\hrule \hbox{\vrule height 0.2 cm \hskip 0.19 cm \vrule height 0.2 cm}\hrule}\,}}

\def\href#1#2{#2}

\makeatletter
\@addtoreset{equation}{section}
\makeatother

\def\appendix#1{
  \addtocounter{section}{1}
  \setcounter{equation}{0}
  \renewcommand{\thesection}{\Alph{section}}
  \section*{Appendix \thesection\protect\indent \parbox[t]{11.15cm}
  {#1} }
  \addcontentsline{toc}{section}{Appendix \thesection\ \ \ #1}
  }

\bigskip
\newcommand{\xx}{\mbox{x}}

\newcommand{\xz}{\hspace{.5pt}\mbox{\sc x}\hspace{.5pt}}

\begin{document}
\renewcommand{\footnotesize}{\small}
\addtolength{\abovedisplayskip}{1.5mm}

\addtolength{\belowdisplayskip}{1.5mm}

\addtolength{\belowdisplayshortskip}{1.5mm}

\newcommand{\Xp}{X^I_+}
\newcommand{\Xm}{X^I_-}
\begin{titlepage}
\hfill
\vbox{
    \halign{#\hfil         \cr
           } 
      }  
\vspace*{35mm}
\begin{center}
{\Large \bf D2 or M2?}\\[1cm]

{\large \bf A Note on Membrane Scattering}

\vspace{1.8truecm}
\centerline{
    {Herman Verlinde}\footnote{verlinde@princeton.edu}}
   \vspace{1cm}
   
\centerline{{\it Department of Physics,
Princeton University}}
\medskip

\centerline{{\it Princeton, NJ 08544, USA}}
\vspace{.5cm}

\vspace*{1cm}
\end{center}

\centerline{\bf ABSTRACT}
\vspace*{10mm}

\noindent
Motivated by a physical interpretation of  its correlation functions as membrane scattering amplitudes, 
we re-address whether the Lorentzian BLG theory  can be
quantized such that it preserves  $SO(8)$  superconformal symmetry. We find that this appears to be possible. While the model correctly reproduces
 protected quantities such as chiral primary amplitudes and the four derivative effective action,  we conclude that, as understood 
at present, it gives a relatively unpractical parametrization of the IR dynamics of M2-branes.

\end{titlepage}

\vskip 1cm

\addtolength{\baselineskip}{.4mm}
\addtolength{\parskip}{.5mm}

\renewcommand{\footnotesize}{\small}

\medskip

\bigskip

\newcommand{\Omeg}{{\mbox{\small $\Omega$}}}

\newcommand{\scy}{{ \sp{\mbox{\sc y}}}}

\newcommand{\scx}{{\mathsf q\smp}}
\newcommand{\scc}{{ \sp\hat{\mbox{\sc c}}}}

\newcommand{\scxs}{{\large \sp{\mbox{\small\sc x}}}}

The groundbreaking work of Bagger, Lambert \cite{Bagger:2006sk}\cite{Bagger:2007vi}, and Gustavsson \cite{Gustavsson:2007vu}
has helped uncover valuable new insight into the structure of the the worldvolume theory of 
coincident M2-branes. This rapid development culminated in the recent 
ABJM formulation of the M2-brane worlvolume theory in terms of an ${\cal N}\! =\! 6$ Chern-Simons theory   \cite{Aharony:2008ug}.  
 
 An earlier proposal for an $SO(8)$ invariant formulation
was made in \cite{Gomis:2008uv}\cite{Benvenuti:2008bt}\cite{Ho:2008ei}, based on a Lorentzian 
three-algebra. This model has the required classical symmetries,
but  has several unresolved problems. 
In particular, the classical theory has ghosts,  $X_\pm$. This feature makes it unclear whether the theory can be quantized in a  way that simultaneously preserves unitarity and
 $SO(8)$ superconformal symmetry.\footnote{The proposal in the first version of \cite{Gomis:2008be} was incomplete, and has been retracted.} Moreover, 
 the ghost-free formulation seems directly equivalent to the non-conformal D2 theory \cite{Bandres:2008kj}\cite{Ezhuthachan:2008ch}\cite{Gomis:2008be}\cite{Honma:2008un}. 
As understood at present, 
Lorentzian model indeed appears to provide
an incomplete description of low energy M2-brane physics  \cite{Gomis:2008vc}. Nonetheless, even with this current
assessment, it is worthwhile to examine the reach and limitations of the model,  
and possibly identify a class of physical questions where it can be useful. 

This short note re-addresses the question whether the Lorentzian model
can be quantized such that its amplitudes are non-trivial, $SO(8)$ invariant, and exhibit
superconformal symmetry.  Relative to earlier discussions, we introduce two new ingredients:

\smallskip

\noindent
(i) To decouple negative norm states, the $X_+$ ghost field needs to satisfy its equations of motion, 
$\del^2 X_+=0.$ To get non-trivial correlation functions, we impose this equation of motion
everywhere except at the locations $z_i$ of the local operators ${\cal O}_i(z_i)$. 
Hence $X_+$ is allowed to develop a pole at the $z_i$
\bea
\label{class}
X_+(y) = \sum_i \sp \frac{\scx_i}{|y-z_i|}\, ,
\eea
with $\scx_i$ some (initially) constant $SO(8)$ vectors. Simultaneously, as indicated in fig 1, we will choose a world volume metric such 
that the region around the points $z_i$ takes the form $S^2 \times \mathbb{R}$.
The Lorentzian 3-algebra theory then describes some generalized  version of 2+1 SYM theory, with
certain $SO(8)$-twisted boundary conditions
and with a position dependent effective coupling $g_{YM}$ proportional to the $\scx_i$.

Via the interpretation of $X_+$ proposed in \cite{Banerjee:2008pd},  as a radial center of mass coordinate of the membrane stack, this prescription makes the correlation functions look 
 like scattering amplitudes of asymptotic multi-membrane~states. 
For fixed values of the $\scx_i$ parameters,  $SO(8)$ superconformal symmetry is broken, in the same way that
any given scattering amplitude breaks the space-time symmetries. To get amplitudes
that can be interpreted as $SO(8)$ SCFT correlation functions, one would need to decouple the $\scx_i$.

\smallskip

\noindent
(ii)   By adding an appropriate ghost action, we show that all negative norm states decouple, provided
that the $\scx_i$ parameters are treated as dynamical variables. We will arrange the ghost action
such that the 
$\scx_i$ in fact must attain their saddle point values. 

Correlation
functions thus take the form of a discrete sum
\bea
\label{semi}
{\cal A} = 
\sum_{X_+^{\rm cl}} \, \Big\langle \prod_i {\cal O}_i(z_i) \Big\rangle_{X_+^{\rm cl}}
\eea
where $\big\langle \ldots \big\rangle_{X^{\rm cl}_+}$ denotes the amplitude computed at $X_+ \! = \! X_+^{\rm cl}$, with $X_+^{\rm cl}$ a saddle point value, at which the correlation function is extremized 
\bea
\label{classdef}
\quad  \frac{\delta\ }{\delta X_+} \Big\langle \prod_i {\cal O}_i(z_i) \Big\rangle_{\strut \mbox{\Large $|$} 
X_+ = X^{\rm cl}_+}  = 0 \, .
\eea
The conformal fixed point value of $X_+$ is evidently one of the saddle points, but a given
correlation function may have other extrema.


\begin{figure}[t]
\begin{center}
\parbox{11.8cm}{

\vspace{1cm}

\begin{flushleft}
\includegraphics[scale=.22]{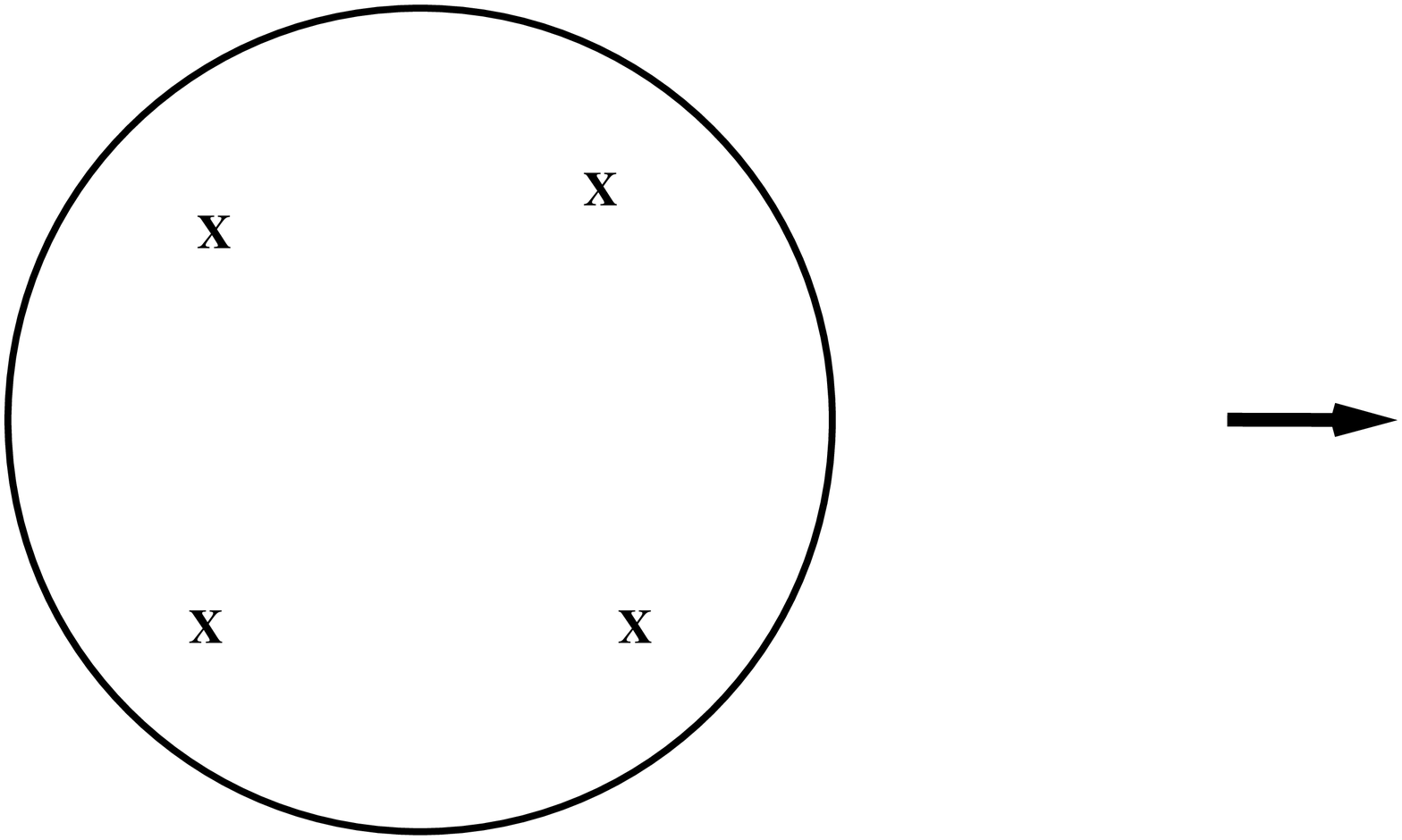}
\end{flushleft}

\vspace{-5.2cm}

\begin{flushright}
\includegraphics[scale=.26]{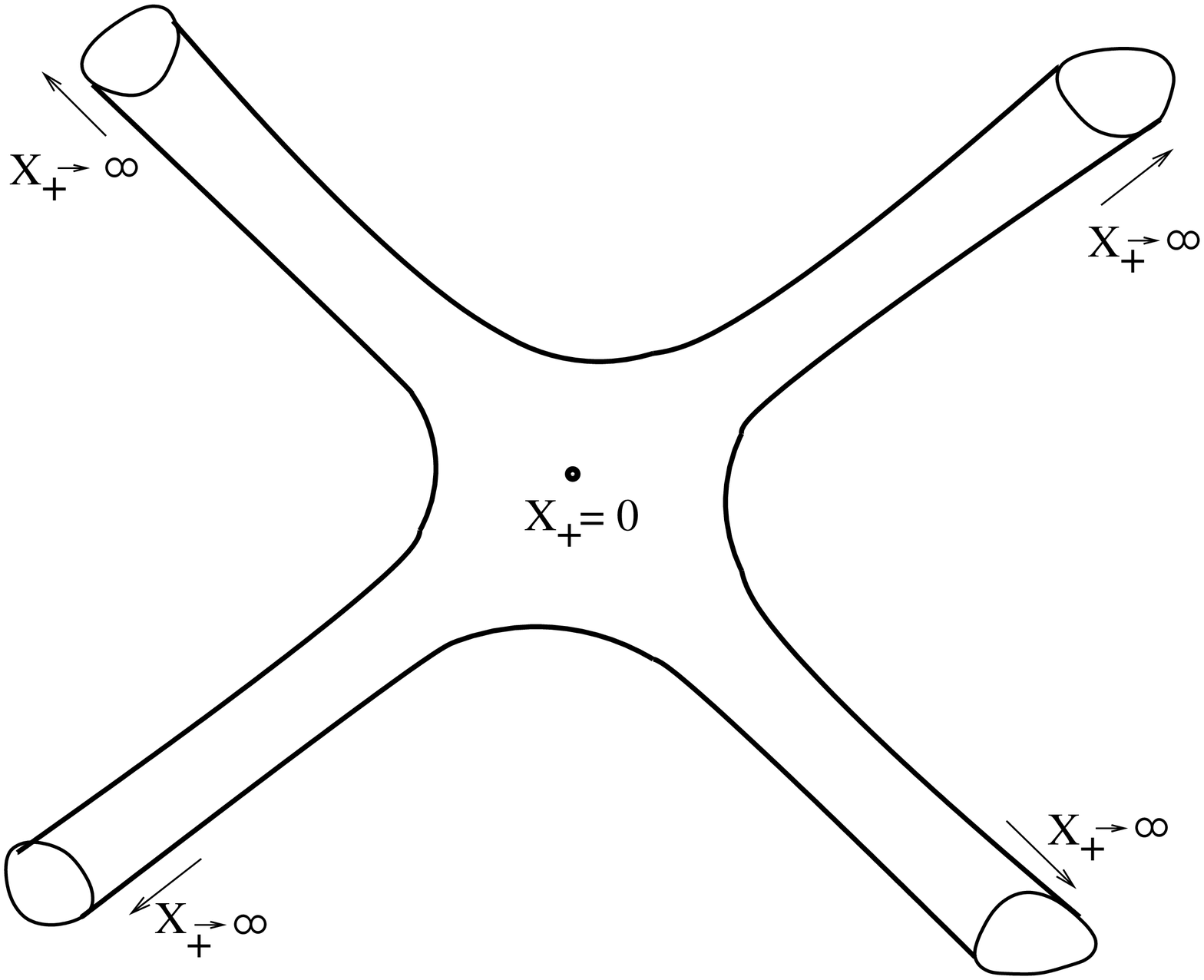}
\end{flushright}}
\caption{$X_+$  is allowed to develop a pole at the location of local operators. Via the proposed identification  $X_+$ as a center of mass coordinate \cite{Banerjee:2008pd},  correlation functions thus
have a suggestive interpretation as scattering amplitudes of asymptotic multi-membrane~states.}
\end{center}
\end{figure}

The proposed formula (\ref{semi})-(\ref{classdef})  has some positive  features:
it does not involve any integral, and formally preserves $SO(8)$ and conformal invariance. It 
nonetheless
appears to have somewhat limited practical use.  The fixed point is expected to 
occur for large  $X_+$, where  the lagrangian becomes strongly coupled. 
Suitably protected quantities, however, such as correlation functions of chiral primary operators and special higher derivative 
terms in the effective action, are still accessible to computation and are correctly reproduced \cite{Paban:1998mp}.  

At a technical level, one could ask whether the prescription of extremizing w.r.t. the parameters $\scx_i$ makes sense. Should we treat them like vevs, and keep them fixed?
Or, since they are determined by local boundary conditions, are we allowed to treat them
as dynamical variables? We return to this question after we have summarized
the formulation that leads to the mentioned results.

\newcommand{\Pssi}{\Psi}

\newcommand{\CC}{\mbox{\small $C$}}
\bigskip

\noindent
{\it Membrane Scattering, Take I}

\noindent
The Lorentzian BLG  action
is a sum 
of kinetic terms (we choose the gauge $D^\mu B_\mu=~0$)
\bea 
\label{classical}
{\cal L}_0 \is {\rm Tr}\Big(\! -\! \frac{1}{2}\big(D_{\mu}X^I \big)^2  + \frac{i}{2}
\bar{\Pssi} \Dslash \sp 
\Pssi + {1 \over 2} \epsilon^{\mu\nu\lambda}
B_{\mu} F_{\nu \lambda} \Big)  \nonumber \\[-1.5mm]
\label{spm}\\[-1.5mm]
{\cal L}_{\rm \pm} \is
 \partial^{\mu}X_- \partial_{\mu}X_+ 
- \frac{i}{2}\Bigl( 
 \bar{\Pssi}_-
 \dslash\sp \Pssi_+ \, - \, h.c.\Big), \nonumber
\eea
and an interaction term 
\bea
{\cal L}_{int} \is  - \frac{1}{2} (X^I_+)^2(B_\mu)^2 
- \frac{1}{12} {\rm Tr}\big(X_+^I [ X^J, X^K] + X^J_+ [ X^K , X^I ] + X_+^K [ X^I , X^J ]\big)^2 \nonumber \\[-2mm]
\\[-2mm]
& &\; +\;  \frac{i}{2}{\rm Tr}\big(\bar{\Psi}\Gamma_{IJ}X_ +^I[X^J,\Psi]\big)\; +\, \frac{i}{4}{\rm Tr}\big(\bar{\Psi}\Gamma_{IJ}[X^I,X^J]\Psi_+\big) +\, h.c.
 \nonumber
\eea
The kinetic term for the ghost fields $X_\pm$ and $\Psi_\pm$ is non-positive definite.

The above lagrangian is $SO(8)$ invariant. For any non-zero value of $X_+$, however, $SO(8)$
invariance is  spontaneously broken: the eight scalar fields $X^I$  split up in seven transverse components
and one longitudinal component in the direction of $X_+$. For constant $X_+$,  the Lorentzian theory reduces to the D2 world volume theory, and the 
transverse components become the seven scalars of 2+1 SYM, and the longitudinal component
is the magnetic dual scalar to the non-abelian gauge field $A$, \cite{Gomis:2008be}. 

The $X_+$ field satisfies the free field
equations 
\be
\label{free}
\del^2X_+ =0.
\ee 
The space of such solutions depends on boundary conditions.
It is therefore useful to contemplate the physical interpretation of  $X_+$.

From the world volume perspective, $X_+$ governs the effective  SYM  coupling constant via
$g_{YM}^2 = \big(X^I_+ \big)^2 .$
To get a new and non-trivially interacting theory, we thus need to choose boundary conditions such that 
$X_+$ is  non-vanishing and non-constant. 
From the target space perspective,  $X_+$ appears to behave as a center of mass 
position of the stack of membranes  \cite{Banerjee:2008pd}, as defined by averaging the locations of all branes within the stack. 
This average can be defined locally on the collective world volume, and thus the center of mass location
may vary with the world volume coordinates. 

With this motiation, we will  allow the $X_+(z)$ field to develop a pole at the location $z_i$ of the local operators as given in eqn (\ref{class}). The coefficient $\scx_i$ of the pole at $z_i$ can be interpreted as the center of mass momentum of the multi-membrane state created by the operator at the location $z_i$. Simultaneously,
we will define  the local operators  at a point $z$ via
`radial quantization': we use the  log of the radial distance to 
the point $z$ as `time', and construct the Hilbert space by 
expanding the fields in spherical harmonics on the $S^2$ surrounding $z$. 
Concretely, this means that, 
for computing correlation functions 
\bea
{\cal A} = \Big\langle \prod_i V_{{\cal O}_i}(z_i, \scx_i)\Big\rangle,
\eea we 
will not choose a standard flat metric, but take it of the  form
\be
\label{metric}
\qquad ds^2 = \sigma^2(y) dy^\mu dy_\mu, \qquad \quad \sigma(y) =  
\sum_i \frac{R}{|y-z_i|} 
\, .
\ee
In this metric,  the worldvolume develops a tube like region
of the form $S^2 \times \mathbb{R}$ near every operator insertion $z_i$.
Here $R$ denotes the radius of the $S^2$.\footnote{We assume that all fields are conformally coupled, so
that the action is Weyl invariant. 
The original and  Weyl rescaled $X_+$ field are related via  
${X}_+ = \sigma \sp \widetilde{X}_+$. Here $X_+$  
satisfies the free eqn of motion (\ref{free}) and may develop poles as in (\ref{class}),  while $\widetilde{X}_+$ remains
finite at  $z_i$.}

To define correlation functions, we need to introduce suitable class of local operators. As a first reasonable guess, we write
the vertex operators that create the asymptotic states in the factorized form
\bea
\label{vertex}
V_{\cal O}(z,\scx) = {\cal O}(z) \,  e^{i\smp \scx \smp \cdot X_-(z)}
\eea
where ${\cal O}(z)$ is a local operator defined out of the transverse fields (i.e. the fields that for fixed and constant $X_+$
constitute 2+1 SYM theory). Examples of such local operators ${\cal O}(z)$ are the chiral primary operators described in \cite{Gomis:2008be}.

The $X_+$-field develops a pole near the vertex operator (\ref{vertex}). 
Since we are using the metric (\ref{metric}), this  does not mean that the
effective  SYM coupling blows up near $z$. To see this, we can
 look at the vertex operator (\ref{vertex}) as a state on $S^2 \times \mathbb{R}$.  Applying  radial quantization  to $X_\pm$ yields a set of creation and annihilation modes $a^\dagger_{\pm, \ell m}$, $a_{\pm,\ell m}$. The $\scx_i$ modes in (\ref{class}) correspond to 
constant zero modes $a_{+,0}$.
The state created by the vertex operator (\ref{vertex}) factorizes as 
\be
|V_{\cal O}\rangle = |{\cal O}\rangle |0,\scx\rangle
\ee
where $|0,\scx\rangle$ is annihilated by all annihilation modes with $\ell>0$, while
for the zero modes 
\bea
\label{qdef}
a_{-,0}\sp |0;\scx \rangle\,  =  0\, ,  
\qquad \qquad 
a_{+,0}\sp |0;\scx \rangle\,  =  \scx\, | 0;\scx \rangle\, .
\eea
In this state, the theory locally reduces to SYM on $S^2 \times \mathbb{R}$ with 
coupling $g_{\rm YM}^2 = \scx^2 /R^2$, or in dimensionless units, $g^2_{\rm YM, eff} = \scx^2/R$. 
Constructing  the quantum operators ${V}_{\cal O}(z,\scx )$ thus still requires control over the interacting 2+1 SYM theory. For BPS operators, this problem appears to be tractable.

All this would seem to yield a satisfactory prescription for computing amplitudes, with the expected
symmetries to support the interpretation as scattering of multi-membrane states in a flat 11-d space-time.
The individual vertex operators (\ref{vertex}) are $SO(8)$ covariant: the $SO(8)$ symmetry
is broken to $SO(7)$ by the direction of the center of mass momentum~$\scx$.
However, there is
the problem of unitarity. The lagrangian (\ref{spm}) has a non-positive definite kinetic term, and while
the specific vertex operators (\ref{vertex}) produce positive norm states,  the above prescription 
does not provide a mechanism for proving that negative norm states can be consistently decoupled from
physical processes.

\bigskip

\bigskip

\noindent
{\it Adding Ghosts}

To eliminate negative norm states,  
 it was proposed in  \cite{Gomis:2008be} that it is sufficient to introduce
a free, non-interacting ghost sector $(\ccc_\pm, \chi_\pm)$. This procedure
 indeed gives rise to a unitary model.
 Ghost fields, however,
should also serve to provide  
a proper integration measure on the classical configuration space, and the minimal set-up of \cite{Gomis:2008be} appears to be insufficient in this respect. Here we will introduce a ghost sector that
not only eliminates negative norm states, but also helps define the $X_+$ integral. In fact, it reduces it
to a sum over saddle points as in (\ref{semi}).

As a quick  guide to the following discussion, here's a toy example of an integral that reduces to a sum over saddle points 
\bea
\label{simple}
{\cal Z} \! \is \! \int \! {dx\smp dx^* d\theta\smp d\theta^*} \,
e^{iS(y)\, + \, i x^* S'(y)} ,
\quad \qquad y = x + \frac{\theta\smp \theta^*}{ x^*} , 
\eea
where $\theta$ and $\theta^*$ are anti-commuting. Performing the $\theta$, $\theta^*$
and $x^*$ integration gives\footnote{Here we used $\int dx^* e^{i x^* p}/x^* = i\pi \sp {\rm sign}(p)$, and dropped a factor of $2\pi i$. The sum is only over minima of $S(x)$.} 
\bea
\label{toy}
{\cal Z} \is   \int \! \! dx\sp \big(\mbox{\footnotesize $\frac{i}{2} |S'(x)| +  S^{''}\!(x) \delta(S'(x))$}\big) \sp e^{iS(x)}
=  2 \sum_{x_{\rm cl}} \, e^{iS(x_{\rm cl})} \, .\nonumber
\eea
The path integral over our ghost sector will have a form very similar to~(\ref{simple}).

With this motivation, we extend the ghost sector 
by `complexifying' all ghost fields $X_\pm$, $\Psi_\pm$, $\ccc_\pm$
and $\chi_\pm$. The complex conjugate fields will be denoted by $X_\pm^*$, etc, but
should be viewed as independent fields from $X_\pm$, etc. The free ghost action reads
\bea
\label{spmtwo}
{\cal L}_{\rm {\pm}} \is \partial^{\mu} X_-^{*}
\partial_{\mu} X_+
- {i} \bar{\Psi}^*_- \dslash 
\Psi_+  
\; +\; h.c.\nonumber \\[-2mm]\\[-2mm]
{\cal L}_{\rm {gh}} \is \,  \partial^{\mu} \ccc_- 
\partial_{\mu} \ccc_+ \; 
- \, {i} \bar{\chi}_- \dslash
\chi_+ \, \; 
+ \; h.c.\nonumber
\end{eqnarray}
and has $SO(8)$ supersymmetry. The $c_\pm$ and $\chi_\pm$ fields have opposite statistics to their counter parts $X_\pm$ and $\Psi_\pm$, and are designed to cancel their quantum fluctuations.

In principle, we could try to derive this ghost sector by applying the rules of BRST quantization to some proper gauged version of the Lorentzian 3-algebra lagrangian.
We will not attempt to do so here. Instead,  we just introduce all the ghost fields by hand, and postulate that the physics needs to be invariant under the  nilpotent BRST transformations
{
\bea
\label{newbrst}
\delta_{{}_Q} X_- \! =  \varepsilon \sp \ccc_- \, ,  
\qquad  \delta_{{}_Q} X_+ 
\! = \varepsilon \ccc^{*}_+ , 
\ & & \  \delta_{{}_Q}
 \Psi_- \! = \varepsilon\sp \chi_-, \qquad  \delta_{{}_Q} {\Psi}_+\!  =  \varepsilon\sp \chi^*_+ , \nonumber\\[-2mm]\\[-2mm]
\delta_{{}_Q} \ccc^{*}_-  \, =  \varepsilon X^{*}_-,  
\qquad \delta_{{}_Q} \ccc_+
\! =  \varepsilon X^{*}_+ , 
\ & & \   \delta_{{}_Q}
\chi_-^*= \varepsilon \Psi_-^* , \qquad  \delta_{{}_Q}  \chi_+ = 
\varepsilon \Psi^*_+ \, . \nonumber
\eea

Defining physical states by the $Q$ cohomology 
eliminates all negative norm states.
To ensure BRST invariance of 
the full theory, however, it is necessary to add an extra interaction term $\Delta {\cal L}_{int}$ to the ghost action.
The new term  should also preserve supersymmetry. These two conditions
are highly restrictive, though still allow for more than one solution. Different solutions 
differ by $Q$ exact terms, and correspond to different choices for the `gauge fixing fermion'.

The toy example (\ref{simple}) suggests that we write the new  interaction lagrangian as
\bea
\label{newint}
{\cal L}_{int}(Y_+,\Upsilon_+) \sp + \sp \big\{ \sp Q\sp ,\sp \ccc_+^{I} P_I \sp + \sp \chi_+^{a}\, S_a \big\} \, ,
\nonumber \\[-1mm]
\\[-1mm]
P_I = \frac{\del {\cal L}}{\del X_+^I} ,  \quad \qquad S_a = \frac{\del {\cal L}}{\del \Psi^a_+}\,. \qquad
\nonumber 
\eea
Here ${\cal L}_{int}$ is the same expression as the old interaction lagrangian, with $X_+$ and $\Psi_+$
replaced by  $Q$ invariant modifications $Y_+$ and $\Upsilon_+$.

The new action continues to be invariant under $SO(8)$ supersymmetry, provided that the new variables form a proper supermultiplet
\bea 
\label{ysusy}
 \delta_{\rm susy} Y^I_+ =  i \bar{\epsilon}\sp \Gamma^I \Upsilon_\pm \, ,\qquad 
\quad \delta_{\rm susy} \Upsilon_+ =  \dslash Y^I_+ 
\Gamma^I \epsilon\,.
\eea
These conditions determine the modified fields $Y_+$ and $\Upsilon_+$, modulo $Q$ exact terms. A minimal solution is to take $Y_+$ of the form 
\bea
\label{ydef}
 Y_+^I  \is X_+^I +   \,  \ccc^{*I}_+  \ccc_+^Kv^K 
, \sp 
\qquad \qquad   v^I\!  =
 \frac{\mbox{\small $X_+^{*I}$}}{\mbox{\small $(X^{*}_+)^2$}}, 
\eea  
and solve for $\Upsilon_+$ using (\ref{ysusy}).
We will not need the explicit expression for  $\Upsilon_+$.
 
 The action (\ref{newint}) with $Y_+$ as in (\ref{ydef}) looks similar to the exponential in (\ref{simple}).
The second, $Q$ exact term in (\ref{newint}) has the standard
form of a gauge fixing term, associated with the gauge fixing conditions $P_I =0$ and
$S_a =0$. These gauge fixing conditions can  be recognized as the saddle point
equations for $X_+$ and $\Psi_+$, and it should therefore be no surprise that in the end, the integral over the ghost
fields reduces to the classical sum (\ref{semi}).
 
  \bigskip 

\bigskip

\noindent
{\it Boundary Conditions at Local Operators}

 With the extended ghost sector and BRST symmetry in hand, we can now revisit the definition
 of local operators, and make sure that all negative norm states decouple.
 Applying  radial quantization  to the ghost sector $(X_\pm, \ccc_\pm)$
gives a set of creation and annihilation modes $a^\dagger_{\pm, \ell m}$, $a_{\pm,\ell m}$,
$c_{\pm,\ell m}^\dagger$ and $c_{\pm,\ell m}$, and similar for the complex conjugate fields. Let us pick some $SO(8)$ direction $\Omega$,
and for now just focus on the modes in this direction. 

The BRST charge, that generates the nilpotent symmetry (\ref{newbrst}), can be expanded as
\bea
\label{qcharge}
Q \is 
 \, c_{-,0} \sp a^{*\dagger}_{+,0}\sp  -\,  c^\dagger_{-,0} \sp a^*_{+,0}\sp \; + \; c^*_{+,0} \sp a^{*\dagger}_{-,0}\sp  -\,  c^{*\dagger}_{+,0} \sp a^*_{-,0}\sp 
\; + \; \ldots 
\nonumber 
\eea
where the ellipsis denote oscillators with $\ell\geq 1$. The  $Q$ cohomology is  trivial: the only obvious
physical ghost sector state is the vacuum $|\sp {\rm 0}  \sp \rangle$
annihilated by all annihilation modes. Via the state operator map, it represents  the identity operator~$1$.

However, there are other possible physical vacua. We can consider vacuum states  
annihilated by all annihilation modes with $\ell\geq 1$ and by all zero modes of the minus fields:
\bea
 a_{-,0} \sp |-1 \rangle\,  = \,0 , \qquad
a^\dagger_{-,0}\sp |-1\rangle\,  =  0\, ,  
\qquad 
c_{-,0}\sp |-1 \rangle\,  = 0\, ,  \qquad  c^\dagger_{-,0}\sp |-1 \rangle\,  = 0\, ,
\eea
and similar for the complex conjugate  zero modes.\footnote{The $|-1\rangle$  vacuum is obtained from $| 0 \rangle$ by a shift in the Fermi and Bose sea level. This type of vacuum
is familiar from superstring theory, where vertex operators are most naturally inserted in the $-1$ picture. The ghost insertion reflect the condition that the
gauge transformations vanish at $z_i$. This restriction creates an extra modulus, that needs to be integrated over. } 
This state  $|\!-\! 1\rangle$ is $Q$ invariant,
but is not annihilated by the annihilation zero modes
of the plus fields.
The path-integral thus includes the integral over the corresponding modes.
We can make this integral explicit via
\be
|-1\rangle = \int \! d\scx\; c_{-,0}^\dagger\sp | 0\sp ,  \scx \rangle \ \times \ h.c.
\ee
where $|0, \scx \rangle$ is the state with given center of mass momentum $\scx$ defined in (\ref{qdef}).

The state $|\!-\! 1\rangle$ is annihilated by all lowering operators of the superconformal algebra, but
unlike the standard vacuum $|\, 0\rangle$, not by all superconformal generators. Instead, it
forms the lowest component of a supermultiplet.

\newcommand{\Omegs}{{\mbox{\tiny\smp$\Omega$}}}

Via the operator state map, $|\! -\! 1\rangle$ corresponds to the local vertex operator
\be
\label{vminus}
V_{-1}(z) = \int \! d\scx \; c_-(z) \sp e^{i \scx \cdot X_-(z)} \ \times \ h.c.\, .
\ee
This operator eliminates the functional integration over
the value of the minus fields $X_-$, $X_-^*$, $\ccc_-$ and $\ccc_-^*$ 
at the location $z$. It
 thereby frees up the integration over the plus modes, that behave as $1/{|y-z|}$. 
 (We'll write these
 modes momentarily.) $V_{-1}(z)$ is not invariant under supersymmetry, but instead forms the lowest component of a superfield. 
 
Local operators at the point $z$ take the form
\be
\label{local}
V_{\cal O}(z) = {\cal O}(z) V^\Omega_{-1}(z),
\ee
where ${\cal O}(z)$ denotes the lowest superfield component of any local gauge invariant operator made up from the transverse modes. In (\ref{local}) we have given $V_{-1}^\Omega(z)$ the superscript
$\Omeg$ to indicate that it involves a particular choice of direction within $\mathbb{R}^8$.
The full
superfield version of $V_{\cal O}(z)$ is given by the product of the superfields of ${\cal O}(z)$ and
$V^{\Omega}_{-1}(z)$. Note that this means that the boundary conditions on the ghost fields depends on
which component of the matter supermultiplet one considers. When defined in this way, the
boundary  conditions at the operator locations preserve supersymmetry.

\bigskip
\bigskip

 \noindent
 {\it Correlation Functions}
 
Correlation functions are obtained inserting the physical operators (\ref{local})
 in the  path integral of the Lorentzian BLG  model, extended with the new ghost sector
\be
\Big\langle \prod_i {\cal O}_i(z_i) \Big\rangle = {\cal N} \, \int {\cal D} \, [{\rm Fields}] \; e^{-S[{\rm Fields}]} \prod_i \, V_{{\cal O}_i}(z_i)  
\ee
with ${\cal N}$ some overall normalization factor, chosen such that $\langle\, 1\, \rangle = 1$. The right-hand side is computed with the metric (\ref{metric}).

The functional integral can be factorized into an integration over the complexified ghost sector times an integral over the transverse modes. 
Since all the ghost fields are non-interacting, it is straightforward
to do their integral. The minus ghost fields are absent from all interactions and
observables, except in the $V_{-1}$ operators. The $V_{-1}$ insertions have the effect of freeing up the
integration over the singular modes 
\bea
\label{singexp}
X_+( y)\!\! \is\!\! \sum_i  \scx_i\,  \mu_i(y) ,\,
\qquad \quad 
\ccc_+( y) =  \sum_i   \scc_i \, \mu_i(y)\, ,\quad  \qquad \ \ \mu_i(y) = \frac{e^{\Omega_i}}{|y - z_i|} \, ,
\eea
and similar  for the complex conjugates fields.
Here $e^{\Omega_i}$ denotes a unit vector in the  $SO(8)$ direction $\Omeg_i$, and indicates the
$SO(8)$ orientation of the modes associated with $\scx_i$ and $\scc_i$.

All regular ghost modes can be integrated out,
producing a trivial overall factor of 1.
What remains is the finite 
dimensional integral over the `moduli', $\scx_i$, $\scc_i$ and their complex conjugates.
\footnote{These parameters indeed play a somewhat similar role as the 
complex structure moduli in the path integral expression for string amplitudes. 
}
This integral looks very similar to the toy example (\ref{simple}), and  it similarly reduces 
to the sum (\ref{semi}) over semi-classical saddle points. This result was anticipated, given that
the last term in the interaction lagrangian (\ref{newint}) can be viewed as a gauge fixing term, that imposes the saddle point equations as a gauge condition.

We thus arrive at the prescription (\ref{semi})-(\ref{classdef}). Assuming
that  correlation functions for fixed $\scx_i$  are finite and reasonably well-behaved,  
it  in principle provides a  concrete answer, that preserves all symmetries of the classical Lorentzian
3-algebra lagrangian, in the sense that they are broken only by the local operators insertions.  Let us briefly address each of the three symmetries -- $SO(8)$ invariance, supersymmetry, and conformal
invariance.

The boundary conditions at the local operators imposed by $V^\Omega_{-1}(z)$ break
$SO(8)$ invariance, since they depend on a choice of direction in $\mathbb{R}^8$.  In the interacting
theory, the operator ${\cal O}(z)$ is sensitive to this choice of orientation. However,
we can restore rotation symmetry,  by including into the definition of the amplitudes, 
the integral over all angles $\Omeg_i$. In the language of 
membrane scattering, this integral amounts to performing an $s$-wave projection
on  all asymptotic states. While this projection removes one apparent source of $SO(8)$ symmetry breaking, 
the true origin of $SO(8)$ invariance is that  2+1 SYM theory is expected
to flow to an $SO(8)$ invariant IR fixed point SCFT (now conjectured to be described, albeit
in a non-manifestly $SO(8)$ invariant way, by the $k=1$ ABJM theory). The RG flow thus
erases, at least locally, the dependence of the local operators on the  orientation $\Omeg_i$.

Supersymmetry is preserved, provided that  one uses  the proper superfield version of the vertex operators, given by the product of the superfield of $V_{-1}(z)$ and the superfield of ${\cal O}(z)$. The  
 $V_{-1}(z)$ superfield is obtained  by replacing $c_-$ and $X_-$ in eqn (\ref{vminus}) by their respective
superfields. Scale invariance follows from the fact that field that
sets the scale, $X_+$, is dynamically driven to attain its saddle point value.\footnote{A simple analogy is
the point particle action, where  minimizing the non-scale invariant action
 $\dot{x}^2/e + m^2 e$ with respect to $e$ gives the scale invariant $m\sqrt{\dot{x}^2}$.}

\bigskip
\bigskip

\noindent
{\it Discussion}

One technical argument one could perhaps raise, is that the moduli $\scx_i$ and $\Omeg_i$ should not be treated as dynamical variables, but represent vacuum expectation values that should be held fixed. However,
as shown above, the value of the moduli represent
properties of a local operator $V_{-1}(z)$, or equivalently, of the
corresponding state $|\! -\! 1\rangle$ defined on $S^2$. In a Hilbert space defined
on a compact spatial slice, one is free to define projection operators.  We conclude therefore that one is allowed to integrate over
the moduli. Moreover, as we have seen, decoupling of negative norm states in our formulation
in fact {\it requires} that we treat the moduli $\scx_i$ as dynamical.

Another more fundamental point of criticism is that the Lorentzian model fails to reach its target because the variables used
in the  lagrangian become strongly coupled for large $X_+$ and thus 
provide an unpractical parametrization of the IR physics of M2-branes.
In view of the high degree of supersymmetry, however, one could still be optimistic that, either 
by some clever insight similar to that of matrix string theory in 1+1 dimensions, or by focusing
on the right type of quantities, one can still retain control over the theory, or  aspects thereof, 
at large values of $X_+$.

An illustrative example of a non-trivial quantity, that can be explicitly computed and has a well-controlled strong coupling limit, is the four derivative term in
the effective action. For 2+1 SYM theory this term has been calculated to all non-perturbative orders
in \cite{Paban:1998mp}. It involves a single perturbative contribution and an elaborate infinite sum over monopole instanton contributions. Via a Poisson resummation, and translated to the Lorentzian 3-algebra model, 
the result can be recast in the schematic form
\be
 \sum_n \frac{\big((\partial X)^2\big)^2 } {\big((X_\perp)^2 + (X^8 + n X_+)^2 \big)^3}
\ee
where $X_\perp$ denote the components of the scalar fields $X^I$ perpendicular and $X^8$ the component parallel to $X_+$. The above expression has a saddle point value at $X_+\! \to\! \infty$, yielding the $SO(8)$ invariant and conformally invariant answer
\be
 \frac{\big((\partial_\mu X)^2\big)^2 } {\big((X )^2 \big)^3}\, .
\ee
This result also matches with the expectations derived from the proposed interpretation of the model as describing M2-branes interacting via 11-d supergravity. Although based on old calculations and derived from 2+1 SYM only, this correspondence still represents a non-trivial test of the M2-brane interpretation of the model.

\bigskip

\noindent
{\it Conclusion}

In this note we have addressed some of the technical and practical criticisms of the
proposed interpretation of the Lorentzian BLG theory as describing the world volume
theory of M2-branes.
We believe that the treatment of the model as presented here answers the main technical objections,
and also highlights more clearly the added structure relative to the pure D2-brane
theory. Nonetheless,  the physical content of the Lorentzian 
model is still very closely linked with that of D2-branes. At a practical level, this rather limits the
amount of new physical and  quantitative insight that one can extract from it.
In its current form, the Lorentzian model clearly does not fully capture the same structure and
dynamical aspects of the IR physics of M2-branes, as encoded in the Bagger-Lambert and ABJM lagrangians. Comparing the two perspectives, however, may still be useful as a potential route towards deriving  $SO(8)$ invariance of the $k=1$ ABJM theory.

\bigskip
\bigskip

\noindent
{\bf Acknowledgments}

 The author thanks Jaume Gomis, Diego Rodr\'{\i}guez-G\'omez, Arkady Tseytlin,
 and Mark Van Raamsdonk for useful critical comments. This work is supported by the National Science Foundation 
under grant PHY-0756966.

\renewcommand{\Large}{\large}

\end{document}